\documentclass[aps,prx,preprint,groupedaddress,a4paper]{revtex4-2}

\usepackage[utf8]{inputenc}
\usepackage[english]{babel}
\usepackage[T1]{fontenc}
\usepackage{graphicx}

\usepackage{caption}
\usepackage{microtype}
\usepackage{booktabs}
\usepackage{mhchem}
\usepackage{siunitx}
\newcommand{\KAppa}[1]{$\qty{#1}{W\,m^{-1}\,K^{-1}}$}
\newcommand{\seebeck}[1]{$\qty{#1}{\mu V\,K^{-1}}$}
\usepackage{xcolor,graphicx}

\makeatletter
 \def\@testdef #1#2#3{%
   \def\reserved@a{#3}\expandafter \ifx \csname #1@#2\endcsname
  \reserved@a  \else
 \typeout{^^Jlabel #2 changed:^^J%
 \meaning\reserved@a^^J%
 \expandafter\meaning\csname #1@#2\endcsname^^J}%
 \@tempswatrue \fi}

\begin{document}

\title{High-Entropy Skutterudites as Thermoelectrics: Potential Synthesizability, Enhanced Stability and Band Convergence via the Cocktail Effect}

\author{Jose~J.~Plata}
\author{Antonio Morales-Altarejos}
\author{Elena~R.~Remesal}
\author{Victor~Posligua}
\author{Antonio M. Márquez}
\email{marquez@us.es}
\affiliation{Departamento de Química Física, Facultad de Química, Universidad de Sevilla, E-41012,~Seville, Spain}

\date{\today}


\begin{abstract}
High entropy materials offer a promising avenue for thermoelectric materials discovery, design, and optimization.
However, the large chemical spaces that need to be explored hamper their development. In this work, a large family of high-entropy skutterudites is explored as promising thermoelectric materials.
Their potential synthesizability is screened and rationalized using the disordered enthalpy-entropy descriptor through high-throughput density functional theory calculations.
In the case of high-entropy skutterudites, the thermodynamic density of states and the entropy gain parameter appear to be key factors for their stabilization. 
Electronic band structure analyses not only show a reduction in the band gap, which enhances carrier concentration and electrical conductivity, but also a band convergence phenomenon for some specific compositions, which is related to the "cocktail effect".
Analyzing atom-projected band structures shows how band convergence is due to the
simultaneous presence of Fe, Ni, and Co in the compound.
The presence of Rh or Ir, while not contributing to this band convergence effect,
can be directly linked to an increase in system's entropy, which enhances
the thermodynamic stability of these materials.
Transport properties are computed for the most promising compositions, and their dynamical, mechanical, and thermal stability are addressed.
Our results demonstrates that these types of compounds open new avenues, not only to enhance thermoelectric efficiency but also to reduce costs by utilizing more abundant elements and also improving their durability.  
\end{abstract}

\maketitle

\section{Introduction}

Over the past decades, we have witnessed a global energy paradigm shift, as fossil fuels are being replaced by more sustainable energy sources. 
Technologies as photovoltaics or batteries are being optimized to provide the energy density and stability needed for large-scale adoption. 
Despite the fact that around 70\% of produced energy is wasted as heat, thermoelectric technologies have not yet become widely adopted in the market, in contrast to other energy solutions.
This is because thermoelectric materials are often costly and possess only moderate energy conversion efficiencies, which has hindered their widespread implementation~\cite{Freer_jmcc_2020}. 
Improving the thermoelectric efficiency of materials is a challenging task. 
The strong coupling between the electronic and thermal transport properties that govern the thermoelectric figure of merit, $zT$, makes this a demanding optimization problem to address~\cite{Toberer_NatRev_ComputationalThermoelectrics_2017}.
Among the most well-established strategies to enhance $zT$ are microstructure~\cite{Qin_nc_2023, Plata_jmca_2023},  band structure~\cite{HeremansSCIENCE2008, Zhang_BandEng_jacs_2012} and phonon~\cite{Li_nm_2018,Qi_nc_2020,Qian_nm_2021,Yang_n_2024, Yao_prb_2024} engineering.
While new materials with high $zT$ have been discovered in recent years~\cite{Han_jmca_2024}, they still fall short of the cost and efficiency requirements needed to be competitive in the market~\cite{Freer_jmcc_2020}.
Therefore, further exploration of novel materials and innovative synthesis techniques are crucial for advancing the field of thermoelectrics, opening the door to their broader application. 
One of the most promising solutions is the use of high-entropy, HE,  materials, which represent a transformative class in Materials Science~\cite{Schweidler_nrm_2024}.

Originally introduced as "high-entropy alloys" in the early 2000s, these multicomponent alloys contain five or more principal elements, each constituting between 5\% and 35\% of the alloy~\cite{hea1, hea2}.
The term "high-entropy" refers to the high configurational entropy of the material, which arises from the multiple principal elements present in similar proportions~\cite{Miracle_am_2017}.
This high entropy contributes to the stabilization of the material's structure, potentially leading to novel properties.

A set of four core effects for HE materials were initially proposed: i) high-entropy, ii) severe-lattice-distortion, iii) sluggish-diffusion, and iv) cocktail effects, corresponding to four aspects of materials: thermodynamics, structure, kinetics, and performance~\cite{Yeh_jom_2013}.
During the last decade, some of these effects have been re-evaluated, and there is controversy regarding whether they can be considered intrinsic characteristics of HE materials. 
For instance, studies suggest that sluggishness is not always present in HE alloys~\cite{Dkabrowa_jac_2019,Mehta_acscs_2020}.
Nevertheless, the promise of HE materials in thermoelectrics lies in their potential to offer a unique balance of properties that can be tuned for specific applications based on some of these four core effects~\cite{Tang_j_2024, Liu_acsel_2025}.

The stabilization of a single phase, often a cubic one, facilitated by the large entropic term, has an important impact on the electronic transport properties~\cite{Luo_jacs_2020}. 
A homogeneous single-phase minimizes carrier scattering and promotes carrier mobility~\cite{Jiang_s_2021}. 
Simultaneously, the high symmetry of a cubic phase enhances band degeneracy, increasing the Seebeck coefficient, $S$~\cite{Li_jacs_2024}.
For instance, in HE GeTe-based materials, a phase transition from a rhombohedral to a cubic symmetry has been reported~\cite{Liu_pnas_2018}.
This new cubic crystal structure enhances band degeneracy and delocalizes carriers. 
The severe-lattice-distortion can significantly reduce the thermal conductivity by introducing phonon scattering centers. 
This effect stems from the partial occupancy of crystallographic sites by atoms with varying atomic radii and masses.
These partially occupied sites can be described as point defects, which act as the primary factors in reducing the phonon relaxation time~\cite{Jiang_s_2022}.
The severe-lattice-distortions also increases diffusion activation energies, reducing diffusion rates of atoms, a phenomenon known as the sluggish diffusion effect.
This effect not only prevents dislocation motion and atom diffusion in junctions but also stabilizes the lattice, specially at high temperature~\cite{Nutor_aem_2020}.
The thermal stability of high entropy materials can be particularly advantageous in thermoelectric applications, as the thermoelectric figure of merit, $zT$, scales with temperature.
Finally, the cocktail effect refers to the synergetic effect arising from the combination of multiple principal elements.
As result, the properties of HE materials are not merely an average of the properties of the individual components but rather a complex interaction that can lead to new and enhanced properties.
The cocktail effect in HE materials highlights the new, vast, and unexplored chemical space that these materials open up in the search for novel thermoelectrics.
This can lead to a broad range of compositions that can potentially substitute critical and expensive elements, thereby reducing costs.

These new chemical spaces also present significant challenges.
The first obstacle that must be overcome is the accurate prediction of the stability and synthesizability of these complex multicomponent materials.
The rational design of HE materials is challenging due to the lack of reliable models for discovering stable element combinations.
Current approaches rely on empirical descriptors, such as the Hume-Rothery rules based on atomic properties like size or electronegativity~\cite{hHume_1935}, but these frequently fail to  predict the formation of these compounds~\cite{Otto_am_2013}.
Alternatively, other descriptors focused solely on mixing enthalpies~\cite{Pitike_cm_2020,Evans_npj_2021} or entropies~\cite{SarkerHarrington_HEC_NCOMMS_2018} also struggle to fully capture the enthalpy-entropy competition in real materials.
Recently, various models have been proposed that combine enthalpic and entropic features to more accurately capture the importance of both factors in the thermodynamic stability of solids~\cite{Divilov_n_2024, Dey_jacs_2024}.

To apply the previous models, the only requirement is to set a reference structure that will be adopted by the high-entropy material.
Among thermoelectric families, some are known to exhibit cubic symmetry, which is the preferred structure for HE materials, such as lead chalcogenides~\cite{Hu_aem_2018,Wang_acsami_2021} or half-Heusler alloys~\cite{Shi_science_2024, Ghosh_j_2024}.
Indeed, there are numerous examples of HE materials based on these families.
However, despite also being cubic, the skutterudite structure has not been thoroughly explored as a prototype for HE thermoelectric materials. 
To the best of our knowledge, there is only one previous study in which Soliman et al. synthesized (CoNiRhIrRu)Sb$_3$ skutterudite nanoparticles for catalytic purposes~\cite{Soliman_cm_2025}.
Synthesizing HE skutterudites not only has the potential to improve the thermoelectric efficiency of their conventional counterparts but also address some of their weaknesses.
The volatile constituents of skutterudites, such as Sb, can jeopardize their long-term stability, especially at high temperatures~\cite{Zelenka_rsca_2019}.
The large resistance to deformation and hardness usually reported in HE materials would potentially limit the atomic diffusion of volatile elements, thus improving their thermal stability~\cite{Li_jmca_2024}.
Additionally, the large chemical space available in HE materials can be explored to identify more sustainable elements, potentially enabling the partial substitution of Co with other more abundant and less expensive metals such as Fe or Ni.

Here, we apply a high-throughput approach to explore the potential synthesizability of HE skutterudites for the first time. 
Compositions where Co is partially substituted by Ni and Fe are screened to identify novel and sustainable high-entropy thermoelectric materials. 

\section{Methods}

The disordered enthalpy–entropy descriptor, DEED,
\begin{equation}
\mathrm{DEED} \equiv \sqrt{\frac{\sigma_{\Omega}^{-1}\left[H_{\mathrm{f}}\right]}{\left\langle\Delta H_{\mathrm{hull}}\right\rangle_{\Omega}}},
\end{equation}
introduced by Divilov et al.~\cite{Divilov_n_2024} is used to evaluate the synthesizability of HE skutterudites
This descriptor incorporates two key parameters:  the entropy gain, $\sigma_{\Omega}^{-1}$, and the  enthalpy cost, ${\left\langle\Delta H_{\mathrm{hull}}\right\rangle_{\Omega}}$.
Both quantities are calculated based on the thermodynamic density of states, $\Omega(E)\delta E$, as defined by Divilov et al.
This property measures the number of configurationally states within [$E$,$E+\delta E$].
The entropy gain parameter, $\sigma_{\Omega}^{-1}$, is calculated as,
\begin{equation}
    \sigma_{\Omega}^{-1}\left\{H_{\mathrm{f}, i}\right\} \equiv \sqrt{\frac{\sum_{i=1}^n g_i\left(H_{\mathrm{f}, i}-\left\langle H_{\mathrm{f}}\right\rangle\right)^2}{\left(\sum_{i=1}^n g_i\right)-1}},
\end{equation}
where, $H_{\mathrm{f},i}$ and $g_i$ are the formation energy and degeneracy of configuration $i$, respectively  and  $\left\langle H_{\mathrm{f}}\right\rangle$ is the average formation energy of all configurations of the ensemble. 
The  enthalpy cost parameter, ${\left\langle\Delta H_{\mathrm{hull}}\right\rangle_{\Omega}}$,  is defined as,
\begin{equation}
    \left\langle\Delta H_{\text {hull }}\right\rangle_{\Omega} \equiv \frac{\sum_{i=1}^n g_i\left(H_{\mathrm{f}, i}-H_{\mathrm{f}, \text{hull}}\right)}{\sum_{i=1}^n g_i},
\end{equation}
where $H_{\mathrm{f}, \text{hull}}$ is the energy of the hull  at the HE stoichiometry. 

To describe the thermodynamic density of states of the high-entropy skutterudites, a 32-atom conventional cell containing 8 cationic sites was built.
All potential configurations and their degeneracies were obtained using the SOD package~\cite{sod}.
For this cell, 112 inequivalent configurations were identified for a HE skutterudite with four different cations at an equimolar ratio.
Convex hulls were built not only considering the binary skutterudites (MSb$_3$) and the high-entropy alloy, but also including all high-symmetry bimetallic (M$^1_{x}$M$^2_{1-x}$Sb$_3$) and trimetallic (M$^1_{x}$M$^2_{y}$M$^3_{1-x-y}$Sb$_3$) skutterudites (M,M$^1$,M$^2$,M$^3$=Fe,Co,Ni,Rh,Pd,Pt,Ir) in which the stoichiometry follows a Zintl electron count of 24 valence electrons per formula unit~\cite{Luo_nc_2015}.
Although disordered binary and ternary skutterudites were not included in the convex-hull calculations, the small energy differences observed between different configurations in the high-entropy compounds suggest that the convex hull's shape and depth are unlikely to change significantly. 

All structures were fully relaxed (atoms and lattice) using first-principles density functional theory calculations performed with the VASP~6.3~\cite{kresse_vasp,vasp_prb1996} program and projector-augmented wave potentials~\cite{PAW}. 
The total energies were computed using the exchange-correlation functional proposed by Perdew et al.~\cite{PBE} with Grimme D3 Van der Waals corrections~\cite{Grimme_jcp_2010}, and the core electrons described by the potentials suggested by Calderon et al~\cite{Calderon_cms_2015}.
A high kinetic energy cutoff of 500~eV and a dense $\Gamma$-centered 4$\times$4$\times$4 Monkhorst-Pack k-point mesh were used to sample the reciprocal space.
The wave function was considered converged when the energy difference between two consecutive electronic steps was less than 10$^{-8}$~eV.
To obtain the optimized conventional unit cell geometry, both the atomic positions and the lattice vectors were fully relaxed until the maximum force component acting on any atom was less than 10$^{-4}$~eV/\AA, employing a supplementary support grid to mitigate the noise in the computed forces.

Interatomic force constants, IFCs, were calculated using the hiPhive package, which combines forces calculated for random atomic distortions in supercells with machine learning regression~\cite{Eriksson_ats_2019}. The forces were calculated in a $2\times 2\times 2$ supercell (256 atoms) using the same setup as for the geometry optimizations. A two-step approach was applied to set the direction and amplitude of the distortion applied to the atoms~\cite{Plata_cm_2022}. A total of 80 distorted supercells were created for training the final model from which IFCs are extracted. Force constants were calculated via multi-linear regression to the DFT forces using least-squares algorithm. IFCs were calculated with cutoffs for the second, third, and fourth-order terms with values of 9.1, 5.5, 2.8~\AA, respectively. The ShengBTE code was used to calculate the lattice thermal conductivity, $\kappa_l$, through the iterative solution of the Boltzmann transport equation (BTE). Scattering times were computed including isotopic and three-phonon scattering. Memory demands and the convergence of $\kappa_l$ with the number of $\bf{q}$-points were balanced using a Gaussian smearing of 0.05 and a dense mesh of $20\times 20\times 20$ $\bf{q}$-points.
Electrical conductivity, $\sigma$, the Seebeck coefficient, $S$, and the electronic contribution to the thermal conductivity, $\kappa_e$, were calculated using the AMSET package \cite{Ganose2021}. This code solves the Boltzmann transport equation using the Onsager coefficients to predict electronic transport properties with the wavefunction from a DFT calculation as the main input. Scattering rates for each temperature, carrier concentration, band, and $\bf{k}$-point are calculated including scattering due to deformation potentials, polar optical phonons, and ionized impurities. Electronic transport properties were calculated using the conventional (32 atoms) and a  mesh of $6\times 6\times 6$ $\bf{k}$-points. Wavefunction coefficients were obtained using the HSE06 functional proposed by Heyd {\it{et al.}} \cite{HSE_2006_JCP}.

On-the-fly machine learning force field generation, as implemented in VASP~\cite{Jinnouchi_2019} has been employed for the Molecular Dynamics (MD) calculations here reported.
A time step of \SI{1}{fs} is used, the plane-wave cutoff energy is kept at \SI{500}{eV}, and the energy convergence threshold of the electronic steps is set to 10$^{-7}$~eV.
An initial training step of \SI{10}{ps} in the isothermal-isobaric ensemble (NPT) of a 2$\times$2$\times$2 supercell has been run for each material (\ce{CoSb3}, \ce{(FeNiCoRh)Sb12}) and temperature (300 and \SI{750}{K}).
In a second, production step, the MD is run for \SI{0.5}{ns} in the canonical ensemble (NVT) allowing further refining of the machine learned force field.
In all cases, the error analysis shows that the mean absolute error (root mean squared error, RMSE) in energies and forces are below 0.97~(0.53)~meV/atom and 77~(13)~meV/\AA, respectively.
In the worst case, 300 out of the \num{500000} steps of these MD runs were full DFT calculations to re-train the machine learned force field, representing $\sim$0.06\% of the MD steps.
This indicates that the machine learned potential has achieved high accuracy in
describing the atomic environments in these materials.
Moreover, full DFT MD calculations have also been run for \SI{10}{ps} with the same setup as above and we compared the mean squared displacements and radial distribution functions with those obtained from the machine learned force field and they are basically identical, reassuring  that the machine learned force field approach used for the longer MD runs is correct.

\section{Results and discussion}

\begin{figure}[ht!]
\includegraphics[width=0.95\columnwidth]{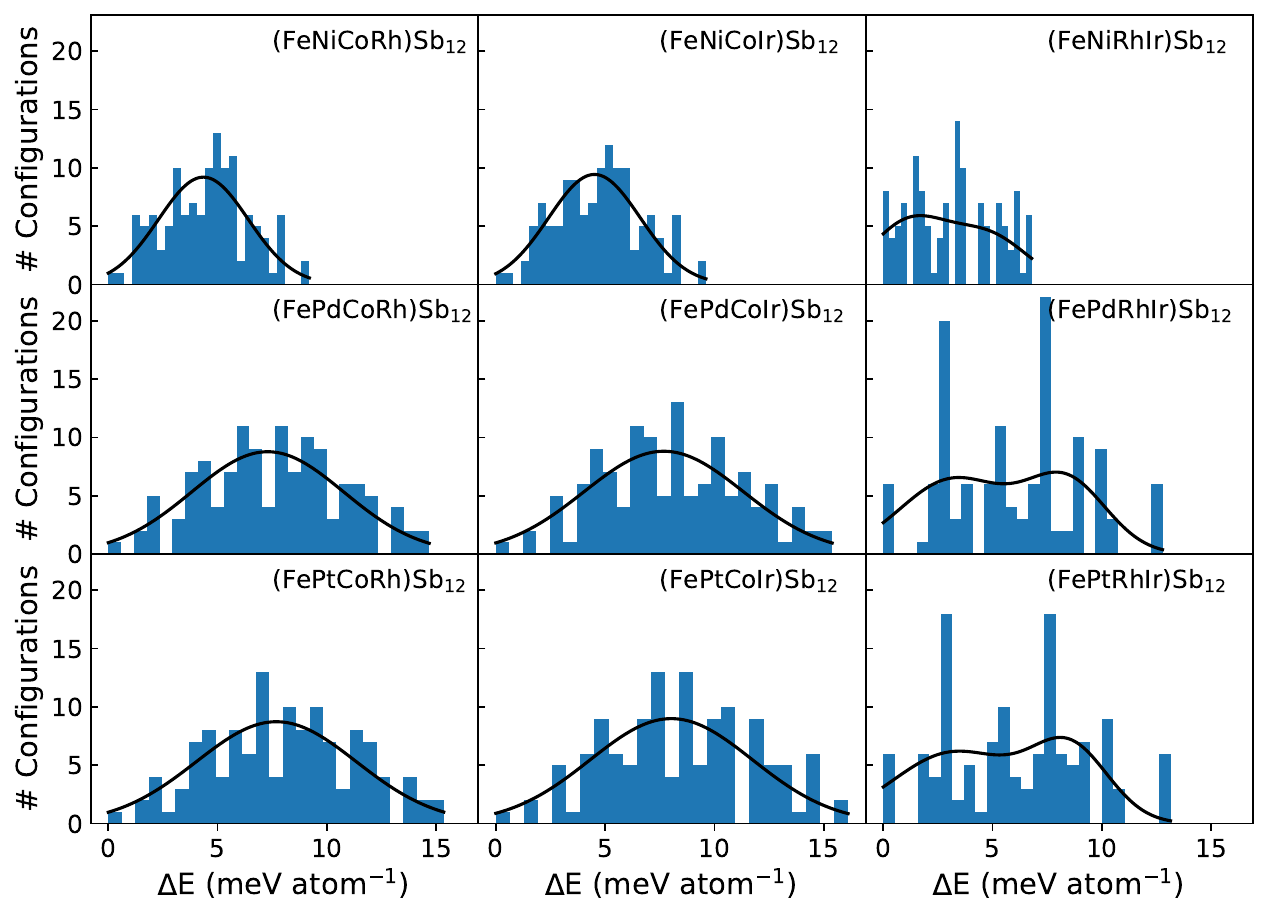}
\caption{Thermodynamic density of states of HE skutterudites studied in this work.}
   \label{fig:tdos}
\end{figure}

{\bf Potential synthesizability}.
In this work nine different HE skutterudites were studied. 
They were obtained combining four elements from groups 8, 9, and 10 with two main conditions: i) Fe should be one of the elements included due to its abundance and low cost, and ii) the number of valence electrons should be 24 per formula unit.
As mentioned earlier, the thermodynamic density of states plays a critical role in the potential synthesizability of high entropy materials.
As a measurement of the thermodynamic density of states, the energy distribution of the 112 inequivalent configurations considering their degeneracy are plotted (Fig.~\ref{fig:tdos}). 
In addition to the entropy gain and enthalpy cost parameters used to compute the DEED, two main features of the thermodynamic density of states are analyzed: bandwidth and configuration distribution.

Firstly, it can be observed that in each row of the panel in Fig.~\ref{fig:tdos}, the range of relative energies or bandwidth, $w$, of each system decreases from left to right: $w_{\ce{FeMRhIr}} < w_{\ce{FeMCoIr}} \simeq w_{\ce{FeMCoRh}}$ (M=Ni, Pd, Pt).
This trend can be explained in terms of the covalent radii (Table~SI-I).
Rh and Ir have much more similar covalent radii to each other than Co, making them more miscible.
In fact, size similarity is one of the variables considered in the classic alloy miscibility rules established by Hume-Rothery.
This greater miscibility is what causes the range of relative energies to be smaller.
On the other hand, when analyzing each column in Fig.~\ref{fig:tdos}, the bandwidth increases from top to bottom.
Therefore, $w_{\ce{FeNiM^1M^2}} < w_{\ce{FePdM^1M^2}} < w_{\ce{FePtM^1M^2}}$ (M$^1$,M$^2$=Co,Rh,Ir).
In this case, this trend cannot be justified solely based on atom size.
The Fe atom, in fact, has a covalent radius whose value is approximately intermediate with respect to the values of Ni and Pd.
However, continuing with the use of the miscibility rules established by Hume-Rothery, the second criterion highlights the importance of having similar electronegativity values (Table~SI-I).
Fe and Ni have much more similar Pauling electronegativity values than the values tabulated for Pd and Pt.
This would explain the greater miscibility and smaller bandwidth for the $\ce{(FeNiM^1M^2)Sb12}$ systems compared to the $\ce{(FePdM^1M^2)Sb12}$ and $\ce{(FePtM^1M^2)Sb12}$.
In addition to  the description of bandwidth, it is important to analyze the distribution of configurations, or equivalently, the thermodynamic density of states across the band. 
Two different trends emerge.
Some systems display a clear maximum in the thermodynamic density of states, typically located at the center of the band.
In these systems, the number of configurations decreases rapidly on either side of the maximum, resulting in a Gaussian-like distribution.
The majority of \ce{(FeMCoIr)Sb12} and \ce{(FeMCoRh)Sb12} (M=Ni, Pd, Pt) systems exhibit this behavior.
Conversely, \ce{(FeMRhIr)Sb12} (M=Ni, Pd, Pt) systems show a flatter distribution, and in some cases, the presence of two maxima is observed.
This may suggest a clustering phenomenon.

\begin{table}[ht!]
    \centering
    \caption{Inverse of the entropy gain, $\sigma_{\Omega}$ parameter, energy above the hull for the ground state configuration, $\Delta H_{\mathrm{hull}}^{\mathrm{GS}}$, the enthalpy cost parameter, ${\left\langle\Delta H_{\mathrm{hull}}\right\rangle_{\Omega}}$, and disordered enthalpy–entropy descriptor, DEED, for HE skuterudites.}
    \begin{tabular}{@{}lccccc@{}}
    \toprule
      Compound & $\sigma_{\Omega}$ & $\Delta H_{\mathrm{hull}}^{\mathrm{GS}}$ & ${\left\langle\Delta H_{\mathrm{hull}}\right\rangle_{\Omega}}$ & DEED \\
      & \multicolumn{3}{c}{(meV\,atom$^{-1}$)} \\
     \midrule
      \ce{(FeNiCoRh)Sb12} & 1.90 & 2.80 & 7.30 & 268 \\
      \ce{(FePdCoRh)Sb12} & 3.15 & 2.10 & 9.72 & 181 \\
      \ce{(FePtCoRh)Sb12} & 3.30 & 1.82 & 9.85 & 175 \\
      \ce{(FeNiCoIr)Sb12} & 1.96 & 5.28 & 10.0 & 226 \\
      \ce{(FePdCoIr)Sb12} & 3.27 & 3.57 & 11.7 & 162 \\
      \ce{(FePtCoIr)Sb12} & 3.42 & 3.24 & 11.7 & 158 \\
      \ce{(FeNiRhIr)Sb12} & 1.99 & 6.41 & 9.62 & 229 \\
      \ce{(FePdRhIr)Sb12} & 3.09 & 3.00 & 9.05 & 189 \\
      \ce{(FePtRhIr)Sb12} & 3.22 & 2.40 & 8.60 & 190 \\
      \bottomrule
    \end{tabular}
    \label{tab:deed}
\end{table}

Convex hulls and thermodynamic density of states were combined to compute the entropy gain,  $\sigma_{\Omega}^{-1}$, and enthalpy cost, ${\left\langle\Delta H_{\mathrm{hull}}\right\rangle_{\Omega}}$, parameters, in addition to the disordered enthalpy-entropy descriptor, DEED (Table~\ref{tab:deed}).
It is important to highlight that small $\sigma_{\Omega}$ values indicate systems that are more stabilized by entropy. 
The systems stabilized by the entropic parameter are those with thermodynamic density of states represented by narrow distributions, with most states being near the central value of the distribution.
The \ce{(FeNiCoRh)Sb12} system exhibits the lowest $\sigma_{\Omega}$, followed by the \ce{(FeNiCoIr)Sb12} and \ce{(FeNiRhIr)Sb12} systems, which is consistent with the previous analysis of the thermodynamic density of states. 
While the $\sigma_{\Omega}$ values varied substantially across the explored compounds, with (FePtCoIr)Sb$_{12}$ having a $\sigma_{\Omega}$ value nearly twice that of (FeNiCoRh)Sb$_{12}$, the enthalpy cost parameter, ${\left\langle\Delta H_{\mathrm{hull}}\right\rangle_{\Omega}}$, shows lower variability.
These values are not only similar but also considerably low. 
For instance, the ground state, GS, structures for each composition are always less than 7~meV\,atom$^{-1}$ above the hull, and the enthalpy cost parameter is also lower than 12~meV\,atom$^{-1}$. 
These values could suggest by themselves potential metastability or synthesizability based on previous analyses comparing experimental polymorphs and $\Delta H_{\mathrm{hull}}$ computed from materials databases such as Materials Project~\cite{Sun_sa_2016,Sykol_sa_2018}
The \ce{(FeNiCoRh)Sb12} system has the lowest enthalpy cost parameter, resulting in the largest DEED among the studied compositions. 
This is followed by the \ce{(FeNiCoIr)Sb12} and \ce{(FeNiRhIr)Sb12} systems, indicating that entropy stabilization plays a prominent role determining the best compositions for their potential synthesis.

Once more likely synthesizable compositions have been determined, theory can also evaluate their potential electronic and thermal transport properties, which is important to screen for new efficient thermoelectric materials. 
In the skutterudite case, lattice thermal conductivity is not the limiting factor for achieving high $zT$ values.
In addition to the severe-lattice-distortion effect  that is well-studied to reduce lattice thermal conductivity in HE materials, empty skutterudites can be filled with "rattle" atoms to further decrease the contribution of phonons to thermal conductivity~\cite{Powell_nature_rc_2025}.
These rattler atoms not only reduce thermal conductivity but also increase the carrier concentration in n-type samples, which, as will be described later, can be beneficial for enhancing the power factor.
That is why, first, we will focus on the electronic structure of these materials as an fingerprint to evaluate their potential use as thermoelectric materials. 

\begin{figure}[ht!]
\includegraphics[width=0.55\columnwidth]{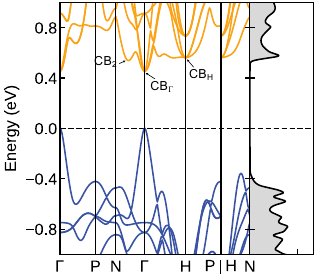}
\caption{Band structure and density of states, DOS, for CoSb$_3$ obtained with r$^2$SCAN functional.}
   \label{fig:cosb3}
\end{figure}

\begin{figure}[ht!]
\includegraphics[width=0.95\columnwidth]{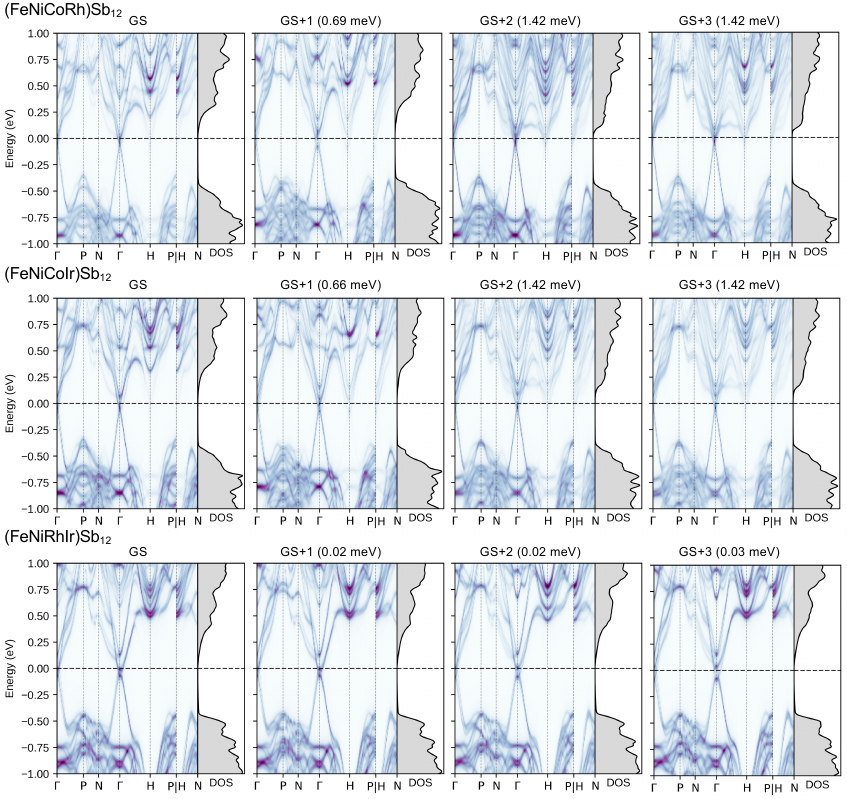}
\caption{Unfolded band structure and density of states, DOS, for HE skutterudites obtained with r$^2$SCAN functional. Most stable configurations are labeled as GS and the other three most stable configurations as GS+X. Energy difference between configurations and GS is included in parenthesis in meV\,atom$^{-1}$.}
   \label{fig:he-fold}
\end{figure}

{\bf Band structure}. The high power factor reported for $n$-type CoSb$_3$ and related materials stems from the contribution of a secondary band, CB$_2$, to the density of states near the Fermi level~\cite{curtarolo:art105}.
The secondary pockets are located in the $\Gamma$-N direction, approximately 0.10-0.15~eV above the conduction band edge at  $\Gamma$, CB$_{\Gamma}$  (Fig.~\ref{fig:cosb3}).
When the Fermi energy is located close to the second band, the Seebeck coefficient ($S$) is drastically increased, primarily due to the enlarged density of states effective mass, $m_d^*$, resulting from the increased number of Fermi pockets, $N_v$~\cite{Lee_acsami_2020}.
This phenomenon is known as band convergence.
The main advantage of band convergence compared to other band engineering strategies, such as band flattening, is that it does not severely affect electrical conductivity.
We have analyzed the electronic band structure of the four most stable configurations for the top three HE compositions identified in the DEED analysis:
\ce{(FeNiCoRh)Sb12}, \ce{(FeNiCoIr)Sb12}, \ce{(FeNiRhIr)Sb12} (Fig.~\ref{fig:he-fold}).

Previous reports indicate that skutterdites band structure is highly sensitive to the exchange-correlation functional, particularly for RhSb$_3$ and IrSb$_3$.
While GGA functionals classify both systems as zero-gap semiconductors~\cite{Koga_prb_2005}, other functionals like the Tran–Blaha modified Becke–Johnson potential predict band gaps exceeding 0.75~eV for both~\cite{Saeed_rscadv_2019}. 
Other meta-GGA functionals, such as r$^2$SCAN or mBJ, have successfully described their band gaps, predicting a symmetry-protected zero-gap semimetal for RhSb$_3$ and semiconducting behavior for IrSb$_3$ (Fig.~S1), aligning with experimental reports~\cite{Wang_prmat_2023}.
Thus, band structures were computed using the r$^2$SCAN functional which balance the accuracy and computational cost~\cite{Furness_jpcl_2020}.
Spin-orbit coupling, SOC,  was also included for some of the structures, but no significant changes were found beyond the breaking of the degeneracy of the three states at the edge of the conduction at $\Gamma$ (Fig.~S1).
To compare the electronic structure of CoSb$_3$ and HE skutterudites, band structure unfolding  was performed using the easyunfold code~\cite{Zhu_joss_2024}.
In all cases, there is a drastic reduction in the band gap, creating conical intersection-like features between the valence and conduction bands.
The zero-gap semimetal behavior of RhSb$_3$ initially suggests that the presence of Rh influences the band gap reduction.
However, this behavior is also observed in HE compositions lacking Rh. 
To understand this trend, the electronic band structure of Fe$_2$Ni$_2$Sb$_{12}$ and Fe$_2$Ni$_2$Co$_4$Sb$_{24}$  has also been computed (Fig.~S2).
In both compositions, the same zero-gap behavior is observed, indicating that the equimolar presence of Fe and Ni is primarily responsible for this band gap reduction.
This band gap reduction will not only increase the carrier concentration but also affect the curvature of the band, indicating a potential increase in electron mobility and, consequently, the electrical conductivity of the material. 
Additionally, the other change in the electronic structure of HE skutterudites compared to \ce{CoSb3} is the appearance of a potential band convergence phenomenon. 
In the case of \ce{(FeNiCoRh)Sb12} and \ce{(FeNiCoIr)Sb12}, in addition to the CB$_2$ previously described in the $\Gamma$-N direction, there are additional pockets that will contribute to conduction due to the band centered at H, CB$_{\mathrm{H}}$.
Although CB$_2$ has a slightly increased energy with respect to CB$_{\Gamma}$ in most explored configurations (0.1-0.12~eV) if compared with the energy difference in \ce{CoSb3} (0.09~eV), the band centered at H is, in most configurations, located at the same energy as the band centered at $\Gamma$.
Band convergence is finally reflected in the density of states, revealing both the energetic interplay between the CB$_2$ and CB$_{\mathrm{H}}$ pockets—characterized by an increase in CB$_2$ pocket energy and a decrease in CB$_{\mathrm{H}}$ pocket energy—and their respective degeneracies ($N_v$). The CB$_2$ pocket exhibits an $N_v$ of 12, while the CB$_{\mathrm{H}}$ pocket has an $N_v$ of 4.
This significantly impacts the DOS: configurations where CB$_2$ is shifted to higher energies show a reduced DOS near the band edge. Conversely, when CB$_{\mathrm{H}}$ convergence occurs, the DOS is higher at the conduction band edge. These modifications will have a deep impact in the Seebeck coefficient and are analyzed in detail in the next subsection.
While this band convergence can be clearly observed in (FeNiCoRh)Sb$_{12}$ and (FeNiCoIr)Sb$_{12}$, the trend is not the same for the (FeNiRhIr)Sb$_{12}$ composition, where CB$_{\mathrm{H}}$ is higher in energy than CB$_{\Gamma}$. 
The band convergence effect that occurs when certain specific elements are combined is what is described as the "cocktail effect" in HE materials.

\begin{figure}[ht!]
\includegraphics[width=0.99\columnwidth]{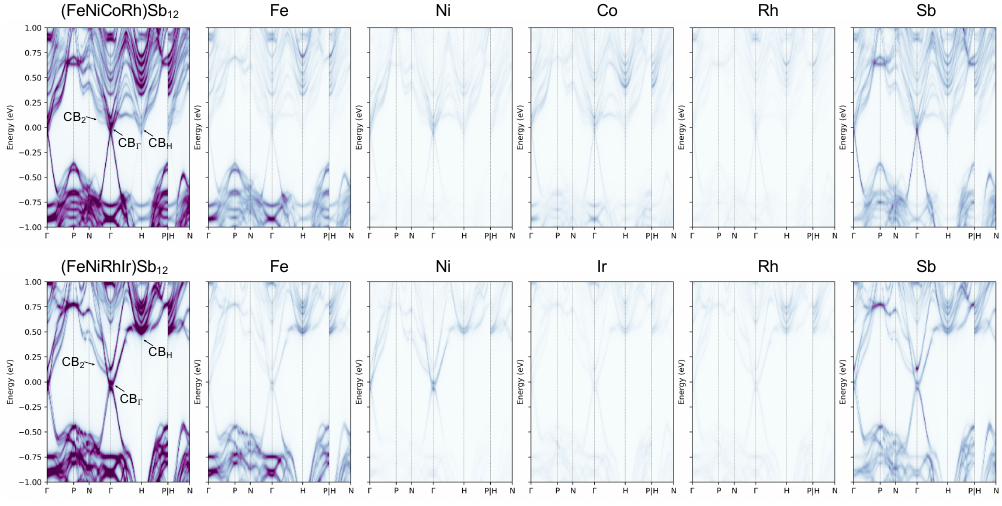}
\caption{Atomic projected band structure of \ce{(FeNiCoRh)Sb12} and \ce{(FeNiRhIr)Sb12}.}
   \label{fig:proyections}
\end{figure}

Analyzing atom-projected band structures can provide valuable insights into the specific atomic orbitals that contribute to the converging bands.
A comparison between the atomic-projected band structures of compounds \ce{(FeNiCoRh)Sb12} and \ce{(FeNiRhIr)Sb12} is included in Fig.~\ref{fig:proyections}. 
The conduction band in both cases shows contributions from all atoms in the compound. 
However, not all atoms participate in the same proportion to all the bands. 
The main contributions to CB$_2$ comes from the Sb-Sb bonds in the Sb$_4$ rings~\cite{Hanus_cm_2017}. 
This band is not as strong in all HE compounds as in the pristine \ce{CoSb3} due to the breakage of symmetry, despite the rings being only slightly distorted in the optimized structures.
Similarly, all metals and Sb contribute to CB$_{\Gamma}$, which has been characterized in previous reports as an antibonding M-Sb state~\cite{Hanus_cm_2017}.
CB$_{\mathrm{H}}$ presents a different scenario.
In compound \ce{(FeNiCoRh)Sb12}, when band convergence is observed, CB$_{\mathrm{H}}$ is formed from the hybridization of Co, Fe, and Ni $3d$ orbitals with Sb $5p$ orbitals.
However, in compound \ce{(FeNiRhIr)Sb12}, CB$_{\mathrm{H}}$ is around \qty{0.5}{eV} higher in energy than CB$_{\Gamma}$, and all four metals and Sb contribute to the band. 
The atomic-projected band structure shows how band convergence is directly linked to the presence of Fe, Ni, and Co simultaneously in the HE compound. 
Meanwhile, although Rh or Ir are not directly participating in CB$_{\mathrm{H}}$, they play a crucial role in increasing the entropy of the system, which favors its thermodynamic stabilization.
Once potential synthesizability and electronic structure have been explored, it is crucial to evaluate the two main factors influencing the applicability of high-entropy skutterudites as thermoelectric materials: stability and thermoelectric performance. 
To do so, we have studied in detail the most four stable configurations of \ce{(FeNiRhCo)Sb12}.
Although measured properties of these high-entropy compounds should be considered as an ensemble of all configurations, the similar  band structure obtained for the most stable configurations suggests that analyzing the lower relative energy configurations provide a reasonable approximation for the properties of the ensemble.

\begin{figure}[ht!]
\includegraphics[width=0.90\columnwidth]{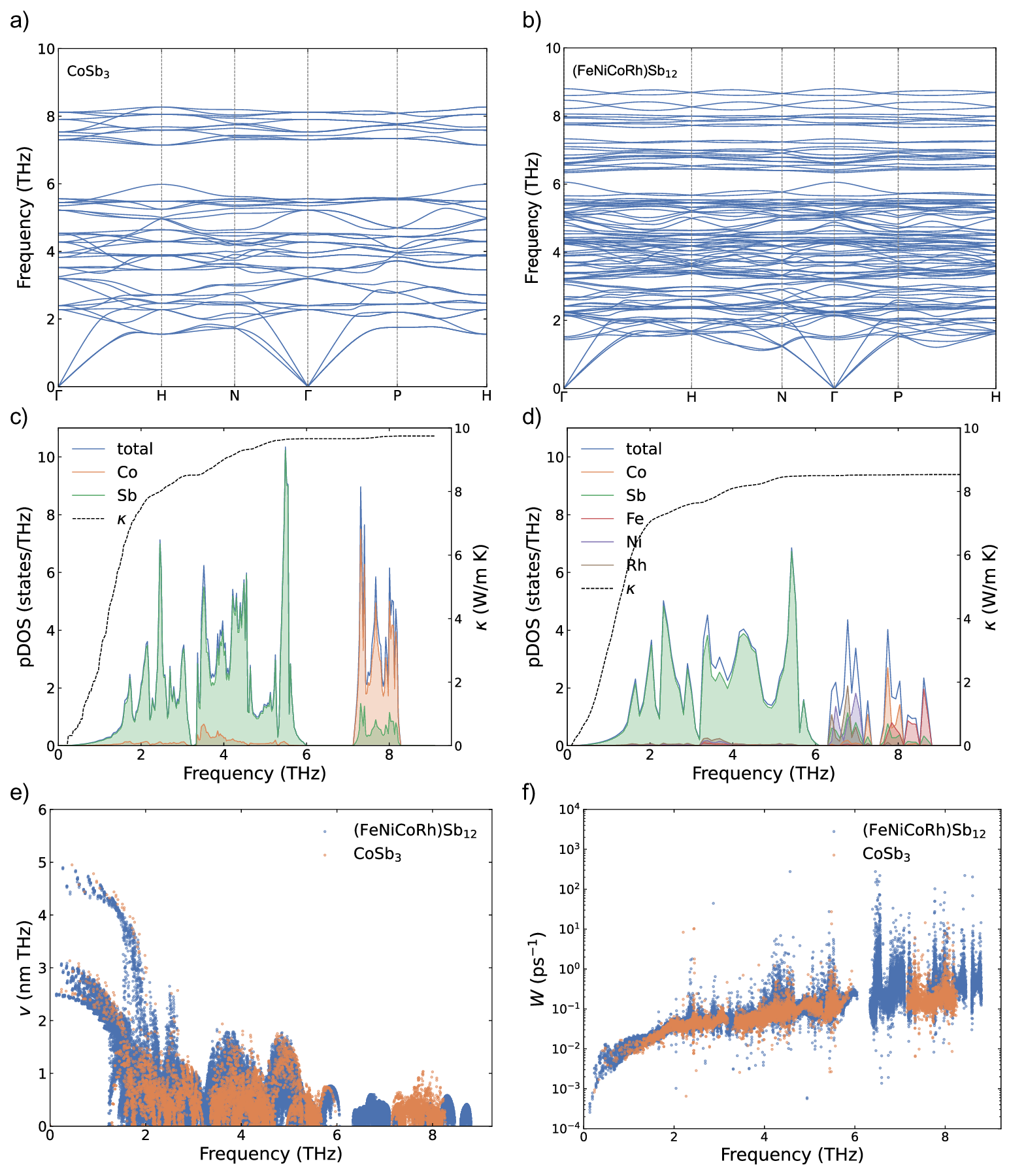}
\caption{a,b) Dispersion curves c,d) vibrational density of states, pDOS, and cumulative lattice thermal conductivity, $\kappa$, e) group velocities, $v$, and f) scattering rates, $W$, for \ce{CoSb3} and \ce{(FeNiCoRh)Sb12} at \qty{300}{K}.}
   \label{fig:phonon}
\end{figure}

{\bf Transport properties}.
In previous studies, the electronic and transport properties of skutterudites have been systematically explored, not only in a wide range of binary compositions but also by screening the effect of pressure in these systems~\cite{Santana_acsami_2024, Lobato_mtp_2025}.
However, applying the same methodology to these systems presents new challenges.
The presence of multiple elements at different crystallographic sites drastically reduces the symmetry of the unit cell, thereby increasing the number of non-equivalent force constants and bands, which leads to a severe increase in computational cost.
This is especially dramatic for thermal conductivity calculations, where a large number of non-equivalent interatomic force constants require the calculation of a larger number of distorted supercells.
Consequently, we have computed the lattice thermal conductivity only for the \ce{(FeNiCoRh)Sb12} ground state.
The calculated thermal conductivity of this high-entropy skutterudite was found to be notably lower than that of its conventional binary counterparts. 
For instance, using the same methodology, \ce{CoSb3}  exhibits a lattice thermal conductivity of \KAppa{9.74} at \qty{300}{K}, whereas the high-entropy \ce{(FeNiCoRh)Sb12} system demonstrates a significantly reduced value of around
\KAppa{8.46}, representing a 13\% reduction.
This reduction is consistent across the entire temperature range; for example, at \qty{800}{K}, \ce{CoSb3} presents a lattice thermal conductivity of \KAppa{3.58}, while the FeNiCoRh high-entropy skutterudite exhibits a lower value of \KAppa{3.09}.
It is important to note that the thermal conductivity of the HE compound is probably overestimated because of the limited size of the model, in which disorder is only explicitly included at short range but not at long range due to the unit cell size.
As an approximation, mass disorder could be included through the Tamura model~\cite{Dominguez_jmca_2025}, obtaining an even lower thermal conductivity of \KAppa{6.9} at \qty{300}{K}.
The reduction in lattice thermal conductivity can be explained by comparing the dispersion curves, group velocities, and scattering rates of \ce{CoSb3} and the \ce{(FeNiCoRh)Sb12} compound. 
Two main differences are observed in the dispersion curves (Fig.~\ref{fig:phonon}a,b): i) low-energy optical modes overlap more effectively with acoustic modes in the HE compound , and ii) high-energy optical modes are split into two bands in the HE system.
While the contribution of high-energy optical modes to thermal conductivity is negligible (Fig.~\ref{fig:phonon}c,d), the effective overlap of low-energy optical and acoustic modes enhances phonon scattering, thereby reducing lattice thermal conductivity.
The primary reduction in lattice thermal conductivity occurs from contributions of the vibrational modes around 1-2~THz, where this band overlap is found (Fig.~\ref{fig:phonon}c,d).
This overlap leads to two effects: i) a slight reduction in group velocities (Fig.~\ref{fig:phonon}e), and ii) a reduction in scattering rates within that frequency range in the HE compound (Fig.~\ref{fig:phonon}f), both of them contributing to a reduction of the lattice thermal conductivity.

\begin{figure}[ht!]
\includegraphics[width=0.99\columnwidth]{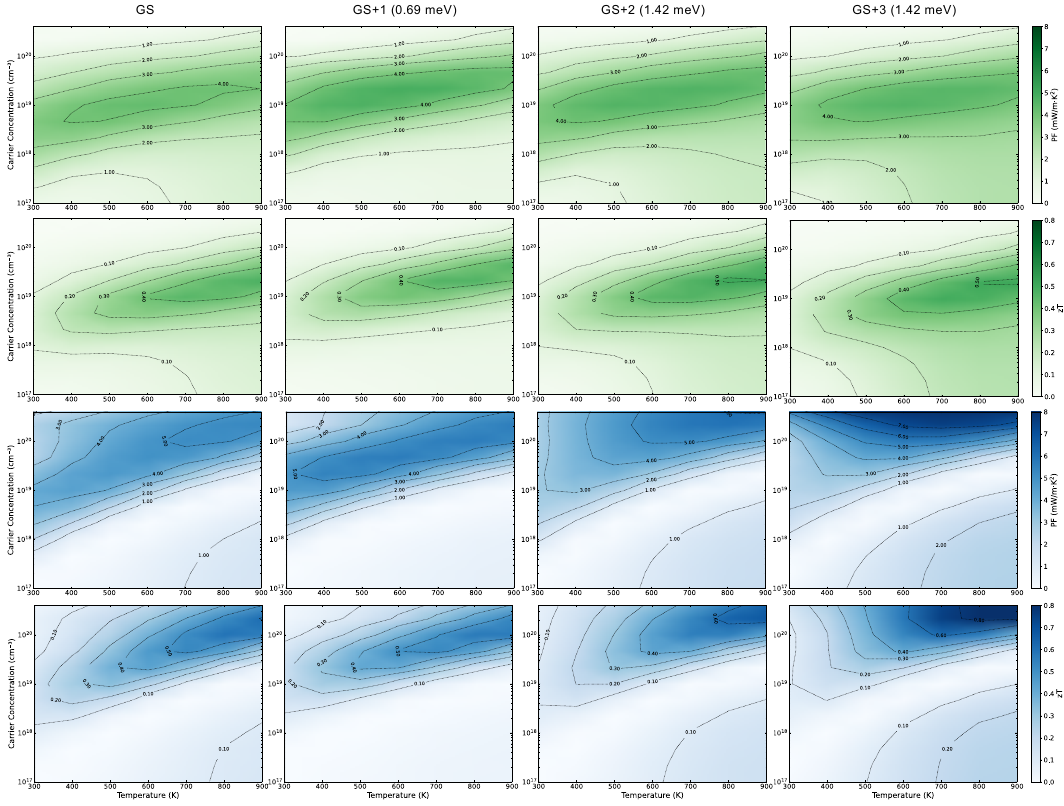}
\caption{Power factor, $PF$, and thermoelectric figure of merit, $zT$, as function of carrier concentration and temperature for \ce{(FeNiCoRh)Sb12} most stable configurations. Green and blue are used for p- and n-type samples, respectively. Energy difference between configurations and GS is included in parenthesis in meV\,atom$^{-1}$.}
   \label{fig:zt}
\end{figure}

While thermal transport properties are strongly bounded to the cationic mass disorder, electronic transport properties are mainly influenced by the band structure.
Although the appearance of a third pocket centered at $H$, CB$_{H}$, would anticipate a significant increase in the Seebeck coefficient, the increase in energy of the secondary pocket, CB$_2$, and its higher degeneracy in all four most stable configurations lead to a lower Seebeck coefficient compared to \ce{CoSb3}.
\ce{CoSb3} single crystals typically exhibit a Seebeck coefficient of approximately \seebeck{373} at \qty{300}{K}~\cite{Caillat_jcg_1996} for a carrier concentration around $10^{19}$~cm$^{-3}$, similar to our predicted values (\seebeck{463}).
In contrast, the HE configurations show values between 151 and \seebeck{228} under the same conditions.
Highest values are observed for the ground state configuration, which slightly preserves the secondary pocket, CB$_2$, close in energy to the main pocket, CB$_{\Gamma}$.
However, in the GS+2 configuration, the secondary pocket is shifted to higher energies, drastically reducing the $S$ to \seebeck{151}.
Nevertheless, given the small band gap, HE compounds are expected to exhibit higher carrier concentrations than \ce{CoSb3}.
If $S$ is compared at a larger carrier concentration, such as $10^{20}$~cm$^{-3}$, the effect of the pocket centered at H becomes apparent.
While Pd-doped \ce{CoSb3} exhibits an $S$ of \seebeck{180} at \qty{300}{K}~\cite{Caillat_jcg_1996} within that carrier concentration range, the GS+3 configuration achieves a value of \seebeck{205}.
This indicates that HE compounds could potentially optimize thermoelectric performance by aligning the Fermi level with beneficial band features.
While the effect of band convergence requires optimization through appropriate carrier concentration, mobilities and conductivities are enhanced in HE compounds due to  a smaller band gap.
For instance, electrical conductivity is almost 5 times larger in any of the four HE configurations studied compared to \ce{CoSb3} at \qty{300}{K} and a carrier concentration of $10^{19}$~cm$^{-3}$.
This high electrical conductivity is not limited to n-type samples but is also observed in their p-type counterparts.
The large electrical conductivity results in a high power factor for both n- and p-type samples (Fig.~\ref{fig:zt}).
Power factors up to 4~mW\,m$^{-1}$\,K$^{-2}$ are obtained for all p-type HE configurations, and 7~mW\,m$^{-1}$\,K$^{-2}$ for the n-type counterpart.
This substantial PF, combined with the low thermal conductivity values, leads to considerably high $zT$ values.
For example, in p-type samples, HE compounds achieve larger $zT$ at lower temperatures.
Using a carrier concentration of $10^{19}$~cm$^{-3}$ as a reference, a $zT$ of 0.3 is obtained for p-type \ce{CoSb3} around \qty{700}{K} (Fig.~SI-3), whereas this value is reached in HE compounds between \qtyrange[range-phrase=~--~]{300}{400}{K}.
In addition to the possibility of achieving higher $zT$ values at the same temperature, this improvement in the figure of merit allows for operation at lower temperatures, which should increase the lifespan of the thermoelectric device.
For n-type samples, a similar trend is observed, yielding $zT$ values around 0.6 at \qty{700}{K}, which could be further increased by reducing lattice thermal conductivity through the use of rattlers. 

\begin{table}[ht!]
    \centering
    \caption{Calculated and experimental elastic constants, $C_{11}$ and $C_{44}$, Young modulus, $E$, Bulk modulus, $B$, Shear modulus, $G$, Poisson's ratio, $\nu$, and Vickers hardness, HV for \ce{CoSb3} and \ce{(FeNiCoRh)Sb12}. Vickers hardness has been computed using microscopic theory of hardness proposed by Tian et al.~\cite{Tian_IJRMHM_Hardness_2012}. Calculated mechanical properties have been obtained using the Voigt-Reuss-Hill average. Values highlighted with * were computed using experimental values.}
    \begin{tabular}{@{}lcccccccc@{}}
    \toprule
      Compound & $C_{11}$ (GPa) & $C_{44}$ (GPa) & $E$ (GPa) & $B$ (GPa) & $G$ (GPa)  & $G/B$ & $\nu$ & HV (GPa) \\
     \midrule
      \ce{(FeNiCoRh)Sb12} & 200 & 54 & 155 & 85 & 65  & 0.76 & 0.20 & 13.0 \\
      \ce{CoSb3} & 182 & 49 & 141 & 85 & 58  & 0.68 & 0.22 & 10.5 \\
      \ce{CoSb3}~\cite{Recknagel_stam_2007} &   &   & 148 & 90 & 61  & 0.67 & 0.23 &  10.9*\\
      \ce{CoSb3}~\cite{Keppens2000} & 158 & 57 & 139*  & 82 & 57  & 0.70 & 0.22*   & 10.7* \\

      \bottomrule
    \end{tabular}
    \label{tab:mech}
\end{table}

{\bf Stability}.
Exploring the stability of a compound requires going beyond an analysis of the thermodynamic stability as it was done before using DEED descriptor, and should also involve assessing its dynamical,  mechanical, and thermal stability, particularly under operando conditions.
Dynamical stability was verified through phonon dispersion calculations, ensuring that no imaginary frequencies exist across the Brillouin zone, which would indicate structural instabilities. 
A  detailed description of the dispersion curves was provided in the previous subsection, highlighting some of the features that governs thermal transport.
Mechanical stability was assessed by calculating elastic constants, confirming that the material satisfies the Born stability criteria, thus ensuring its structural integrity under external stresses, see Table~\ref{tab:mech}. 
For any thermoelectric material application, mechanical properties are as crucial as their thermoelectric counterparts. 
A long-term, reliable thermoelectric device requires high elastic moduli to withstand external forces such as bending or shaking, thereby preventing crack formation or propagation.
The calculated elastic constants for \ce{CoSb3} and \ce{(FeNiCoRh)Sb12} were used to compute most common mechanical properties, showing good agreement with available experimental data \ce{CoSb3}. 
The high-entropy structure exhibits larger Shear and Young's moduli than CoSb$_3$, as well as an approximately 30\% higher Vickers Hardness, indicating greater mechanical reliability.
Thermal stability was assessed through molecular dynamics simulations using machine learning force fields to evaluate the structural integrity and phase stability at 300~K and 750~K.
This ensures their suitability at the temperature where this family of skutterudites is reported to exhibit larger thermoelectric efficiency.
Mean squared displacement, MSD, curves confirmed the structural stability of the materials at both temperatures, showing no signs of atomic diffusion or phase decomposition (Fig.~SI-4).
The average MSD values for both \ce{CoSb3} and \ce{(FeNiCoRh)Sb12} were similar and small: $\numrange[range-phrase=-]{0.029}{0.031}$~\AA$^2$ at \qty{300}{K} and
$\numrange[range-phrase=-]{0.089}{0.092}$~\AA$^2$ at \qty{750}{K}, indicating that the atoms vibrate around their equilibrium positions.
As expected, the vibrational amplitude was higher at the elevated temperature, but the slope of the MSD curves remained zero.
To understand the effect of temperature on the structural motifs of the skutterudite, we performed a detailed analysis of the radial distribution function and the bond-angle distribution.
No changes were observed in the radial distribution functions (Fig.~SI-5) over the first and last \qty{10}{ps} of the simulation. 
At \qty{700}{K}, as anticipated, the first two peaks, corresponding respectively to the metal-Sb and Sb-Sb first neighbors, were wider than at \qty{300}{K}, yet they remained well defined.
In the case of \ce{(FeNiCoRh)Sb12}, the metal-Sb peak around \qty{2.6}{\angstrom} was slightly wider than for \ce{CoSb3}.
This broadening is attributed to the presence of Fe-Sb, Ni-Sb, Co-Sb, and Rh-Sb first-neighbor distances within that peak, which are slightly different from each other.
The stability of the Sb-rings is also important for the material's electronic properties.
The Sb-Sb-Sb angles in the Sb-rings exhibited well-defined distributions for both materials and temperatures (Fig.~SI-6).
In all cases, the observed distributions correspond to a Gaussian profile centered at \qty{90}{\degree} with full width at half maximum of 5.4-\qty{5.9}{\degree} at \qty{300}{K} and 6.9-\qty{9.2}{\degree} at \qty{750}{K}.
No distortion of the \ce{Sb4} rectangles is thus observed.

\section{Conclusions}

In summary, the potential synthesizability of high-entropy skutterudites has been explored for their potential application in thermoelectric devices.
The inherent properties of high-entropy alloys, combined with the crystal structure and large chemical space offered by skutterudites, open the door to the discovery of highly efficient, cost-effective, and durable thermoelectric materials. 
By combining high-throughput DFT calculations with the disordered enthalpy-entropy descriptor, the most likely compositions to be synthesized have been identified.
It has been demonstrated that their stability is primarily described by their thermodynamic density of states, and classic Hume-Rothery rules can be used to rationalize their stability. 
Electronic band structures have been calculated as fingerprints of their electronic transport properties. 
A systematic band gap reduction, which enhances carrier concentration and electrical conductivity, has been reported. 
Notably, a band convergence phenomenon has been observed for certain compositions, indicating the potential to enhance the power factor of these systems. 
This is a consequence of the "cocktail effect" that arises when specific elements are combined in high-entropy materials.
Analyzing atom-projected band structures shows how this band convergence is
due to the simultaneous presence of Fe, Ni, and Co in the compound.
While Rh or Ir are not directly involved in this phenomenon, they play
a crucial role in increasing the system's entropy, which favors its
thermodynamic stabilization.
The calculated transport properties show that while the CB$_\mathrm{H}$ could potentially enhance $S$, the shift of the secondary pocket, CB$_2$, to higher energies, and especially its higher degeneracy, can counteract this phenomenon, particularly at moderate carrier concentrations. 
Nevertheless, the high mobility and electrical conductivity lead to a high power factor, especially at lower temperatures when compared to CoSb$_3$.
Finally, the stability of these compounds has been addressed, demonstrating higher Young's and shear moduli and hardness, which indicates an improvement in durability compared to \ce{CoSb3}.

 
\section{Acknowledgements}
This work was funded by
grant PID2022-138063OB-I00 funded by MICIU/AEI/10.13039/ 501100011033 and by FEDER, UE, and by
grant TED2021-130874B-I00 funded by MICIU/AEI/10.13039/501100011033 and by the “European Union NextGenerationEU/PRTR.
We thankfully acknowledge the computer resources at Lusitania
(Cenits-COMPUTAEX), Red Española de Supercomputación, RES
(QHS-2024-1-0022 and QHS-2024-2-0020) and Albaicín (Centro de Servicios de
Informática y Redes de Comunicaciones - CSIRC, Universidad de
Granada).



\bibliographystyle{apsrev4-2}
\bibliography{xjose}
\includegraphics[width=\linewidth,page=1]{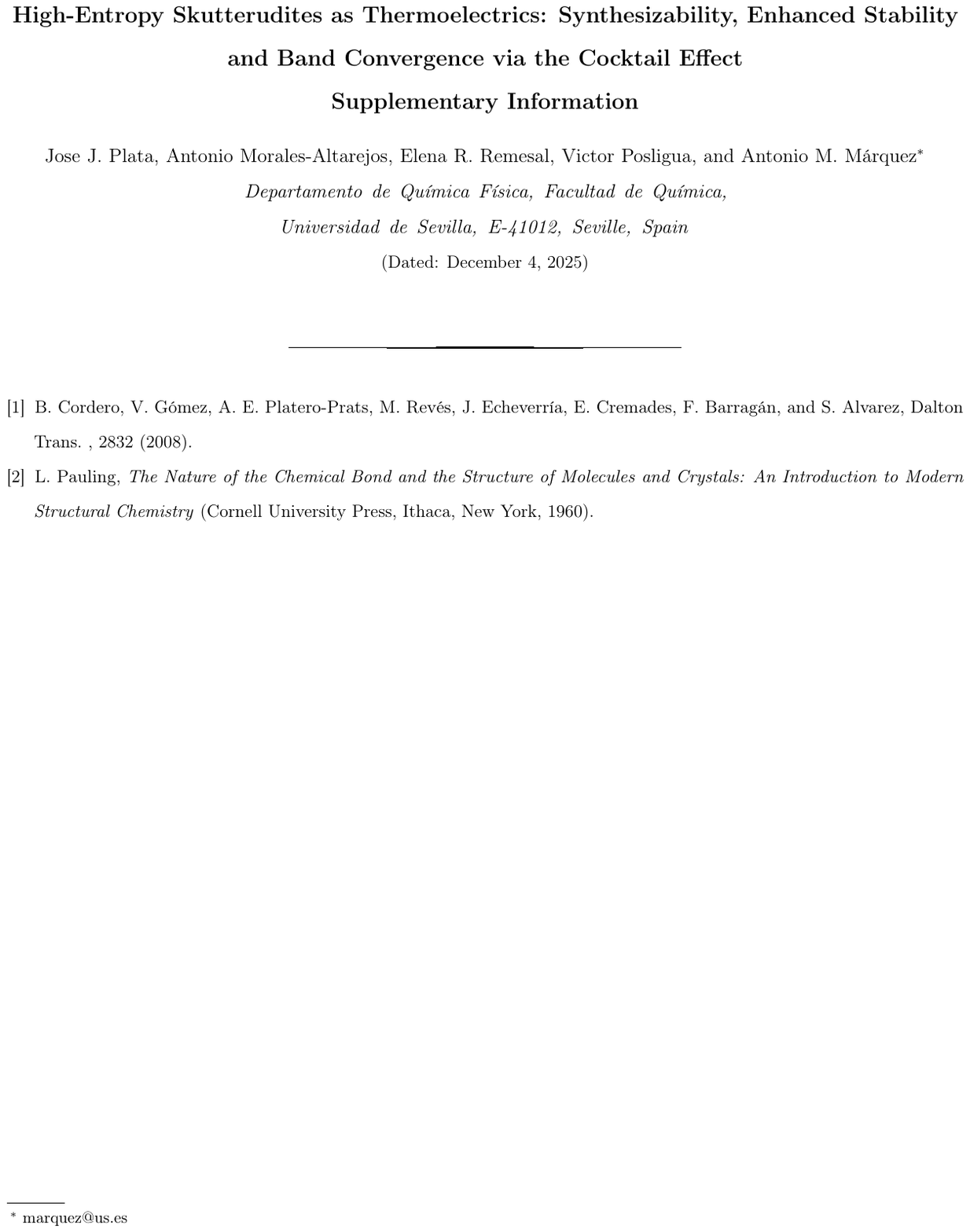}
\includegraphics[width=\linewidth,page=2]{SI.pdf}
\includegraphics[width=\linewidth,page=3]{SI.pdf}
\includegraphics[width=\linewidth,page=4]{SI.pdf}
\includegraphics[width=\linewidth,page=5]{SI.pdf}
\includegraphics[width=\linewidth,page=6]{SI.pdf}
\end{document}